\documentclass[%
 aip,
 jcp,
 amsmath,amssymb,
 reprint,%
floatfix,
]{revtex4-2}

\usepackage{graphicx}
\usepackage{dcolumn}
\usepackage{bm}
\usepackage[T1]{fontenc}
\usepackage[utf8]{inputenc}
\usepackage{physics}
\usepackage{xcolor}
\usepackage[inline]{enumitem}
\usepackage[normalem]{ulem}
\begin{document}

\title{Phase diagram of the hard-sphere potential model in three and four dimensions using a pseudo-hard-sphere potential}

\author{Edwin A. Bedolla-Montiel}
\email{e.a.bedollamontiel@uu.nl}
 \affiliation{Soft Condensed Matter \& Biophysics, Debye Institute for Nanomaterials Science, Utrecht University, Princetonplein 1, 3584 CC Utrecht, Netherlands.}

\author{Ramón A. Castañeda-Cerdán}
\email{ramon.castaneda@cinvestav.mx}
\affiliation{Departamento de Física, Cinvestav, Av. IPN 2508, Col.San Pedro Zacatenco, Gustavo A.\ Madero, 07360,Ciudad de México, Mexico
}%

\author{Ramón Castañeda-Priego}
 \email{ramoncp@fisica.ugto.mx}
\affiliation{
Departamento de Ingeniería Física, División de Ciencias e Ingenierías, Campus León, Universidad de Guanajuato, Loma del Bosque 103, Col. Lomas del Campestre, 37150, León, Guanajuato, Mexico
}%

\date{\today}

\begin{abstract}
The hard-sphere potential has become a cornerstone in the study of both molecular and complex fluids. Despite its mathematical simplicity, its implementation in fixed time-step molecular simulations remains a formidable challenge due to the discontinuity at contact. To avoid the issues associated with the ill-defined force at contact, a continuous potential has recently been proposed - here referred to as the pseudo-hard-sphere potential (pHS) [J. Chem, Phys. \textbf{149}, 164907 (2018)]. This potential is constructed to match the second virial coefficient of the hard-sphere potential and is expected to mimic its thermodynamic properties. However, this hypothesis has only been partially validated within the fluid region of the phase diagram for hard-sphere dispersions in two and three dimensions.
In this contribution, we examine the ability of the continuous pHS potential to reproduce the equation of state of a hard-sphere fluid, not only in the fluid phase but also across the fluid-solid coexistence region. Our focus is primarily on the phase diagram of hard-sphere systems in three and four dimensions, however, we also report on the feasibility of the pHS to reproduce the long time dynamics of a three-dimensional colloidal dispersions. We compare the thermodynamic properties obtained from Brownian dynamics simulations of the pHS potential with those derived from refined event-driven simulations of the corresponding hard-sphere potential. Furthermore, we provide a comparative analysis with theoretical equations of state based on both mean-field and integral equation approximations.
\end{abstract}

\maketitle

\section{\label{sec:intro}Introduction}
The hard-sphere (HS) model remains the reference system for liquids and soft matter due to its simplicity, physical relevance, and rich behavior. It has been extensively studied over the years~\cite{mulero2008theory,hansen2013theory,royallColloidalHardSpheres2024} through computer simulations~\cite{alder1957phase,frenkel2023understanding} and is commonly employed as a reference system in perturbation-based thermodynamic approximations~\cite{scholl-paschingerVaporliquidEquilibriumCritical2005a,valadez-perezPhaseBehaviorColloids2012}. The continued interest in the HS model stems from the extensive knowledge available and the feasibility of experimental realizations~\cite{puseyPhaseBehaviourConcentrated1986,thorneyworkTwoDimensionalMeltingColloidal2017,thorneyworkSelfdiffusionTwodimensionalBinary2017,thorneyworkStructureFactorsTwodimensional2018}, made possible by recent advances in microscopy and scattering techniques. These developments, together with complementary computer simulations~\cite{pieprzykThermodynamicDynamicalProperties2019a} and theoretical studies using integral equation theory~\cite{henderson1976ornstein,santosStructuralThermodynamicProperties2020b}, establish the HS model as a robust framework for testing theories across the full spectrum of current soft matter physics research.

Computer simulations are among the principal techniques in soft matter research and have been extensively employed to study the phase diagram of the HS model~\cite{speedyPressureEntropyHardsphere1998,pieprzykThermodynamicDynamicalProperties2019a,royallColloidalHardSpheres2024}. However, the inherent discontinuity of the HS model renders continuous time integration methods unsuitable~\cite{allen2017computer,frenkel2023understanding}. In particular, Brownian dynamics simulations~\cite{ermakBrownianDynamicsHydrodynamic1978,hansen2013theory} cannot be directly applied to the HS model, necessitating the use of specialized event-driven algorithms~\cite{scala2007event}.
A mapping of the HS model to a continuous interaction potential was proposed by \textcite{joverPseudoHardspherePotential2012}, where
the compressibility factor was compared against the well-established Carnahan-Starling (CS) equation of state~\cite{hansen2013theory}, and the parameters of the potential were fixed when the compressibility factor from simulations matched the analytical CS expression. Rather than relying on trial and error, the work of \textcite{baez2018using} employed the extended law of corresponding states~\cite{noro2000extended} to determine a set of potential parameters by directly matching the second virial coefficient of the HS potential with that of the continuous interaction potential. This procedure yields a pseudo hard-sphere (pHS) interaction potential with parameters that are independent of thermodynamic conditions, such as density.

The physical criterion proposed by \textcite{baez2018using} points toward a deeper understanding of the role of the second virial coefficient in explaining the global and local properties of both molecular liquids and soft materials, an aspect that cannot be conceived a priori and therefore it is important to systematically test it. So far, the mapping of the HS model to a continuous potential has been shown to reproduce the thermodynamics of the real HS one-component fluid~\cite{joverPseudoHardspherePotential2012}, as well as those of binary and polydisperse mixtures~\cite{martinez-riveraModelingStructureThermodynamics2023}. The pHS model has recently been employed to inverse design the self-assembly of exotic crystalline structures~\cite{bedolla-montielInverseDesignCrystals2024}. In addition, computer simulations have been used to investigate the dynamics and transport phenomena of the pHS model~\cite{nicasio2023dynamical,nicasio2023pseudo}, and the interaction potential has found applications in modeling active matter~\cite{torres2024motility} and in evaluating depletion forces in binary and ternary colloidal mixtures~\cite{DelosSantos2022,DelosSantos2024}. These examples underscore the robustness of the model and its wide range of applications in systems where short-ranged, hard-like interactions are required. Furthermore, the pHS interaction potential has been successfully applied within the Ornstein-Zernike framework to reproduce thermodynamic properties, particularly in studies of polydisperse fluids~\cite{martinez-riveraModelingStructureThermodynamics2023}. All these cases make it obvious that, similar to quantum mechanics, there is a group of "isospectral" potentials that lead to identical physical properties. This information is useful not only for proposing a soft and continuous potential that can be used in computer simulation techniques, it highlights the fact that we should explore in more detail the physics that can be extracted from those many-body systems that have the same value of the second virial coefficient.

The aim of this work is to demonstrate the applicability of the pHS model across the full range of the three-dimensional fluid and solid phases and to determine the co-existence densities at the fluid-solid transition using Brownian dynamics simulations. We also provide a brief discussion on the suitability of the pHS potential to replicate the long time dynamics of three-dimensional colloidal dispersions. We solve the Ornstein-Zernike equation with various closure relations to obtain the liquid equation of state over a range of densities. Furthermore, we test the hypothesis proposed by \textcite{baez2018using} that the potential is applicable in arbitrary spatial dimensions by performing computer simulations and solving the Ornstein-Zernike in four dimensions. Our results show excellent agreement with mean-field equations of state, previously reported simulation data used for fitting semi-phenomenological equations of state, and solutions of the Ornstein-Zernike equation.

Following this Introduction, the manuscript is organized as follows. In Sec.~\ref{model}, we introduce the pHS model, outline its derivation, and specify the parameters used for the interaction potential throughout this work. Sec.~\ref{sec:methods} provides a detailed description of the methods employed for the computer simulations and the numerical solution of the Ornstein-Zernike equation, including the relevant parameters. Sec.~\ref{sec:integral} is dedicated to the presentation and discussion of the results obtained from the Ornstein-Zernike equation, focusing mainly on the fluid region. The results of the long-time self-diffusion coefficient as a function of the packing fraction of a three-dimensional colloidal dispersion are discussed in Sec. \ref{sec:diff}. In Sec.~\ref{sec:sims}, we show and examine the computer simulation results for the three- and four-dimensional systems. Finally, Sec.~\ref{conclusions} concludes the work with an overview of the main findings and perspectives based on the results presented.

\section{\label{model}Pseudo-hard sphere potential revisited}

The HS system consists of spheres of diameter $\sigma$ that cannot overlap, which interact via the potential \cite{royallColloidalHardSpheres2024},
\begin{equation}
    u_{HS}(r) =  \left\{
    \begin{array}{cc}
        \infty & r < \sigma, \\
        0 & r \geq \sigma,
        \label{eq:hs}
    \end{array} \right.
\end{equation}
where $r$ is the distance between the spheres centers.

The work of~\textcite{joverPseudoHardspherePotential2012} introduced a continuous interaction potential, which is a generalized cut-and-shifted Mie potential, to reproduce the CS equation of state. The pHS interaction potential is defined as \cite{joverPseudoHardspherePotential2012}
\begin{equation}
    u_{pHS}(r) =  \left\{
    \begin{array}{cc}
        A \varepsilon \left[ \left( \frac{\sigma}{r} \right)^{\lambda}- \left(\frac{\sigma}{r} \right)^{\lambda-1} \right] + \varepsilon & r < \sigma B, \\
        0 & r \geq \sigma B ,
        \label{eq:wca}
    \end{array} \right.
\end{equation}
where
\begin{equation}
    A = \lambda \left( \frac{\lambda}{\lambda -1} \right)^{\lambda-1} \quad , \quad B = \frac{\lambda}{\lambda -1}
    \label{eq:wca_params}
\end{equation}

The exponent $\lambda$ is related to the stiffness of the interaction potential, and $\epsilon$ is the energy parameter that measures the repulsion strength between a pair of particles. Typically, $\epsilon$ is used to define the reduced temperature $k_B T/\epsilon$, where $k_B$ is the Boltzmann constant and $T$ the absolute temperature.

Later, in the work of~\textcite{baez2018using}, it was shown that following the so-called extended law of corresponding states~\cite{noro2000extended}, the assumption that the second virial coefficient of the continuous potential $B_2^{\mathrm{pHS}}$ must be equal to the second virial coefficient of the HS interaction potential, $B_2^{\mathrm{HS}}$, will yield a set of values for $\epsilon/k_B T$ in terms of $\lambda$ that successfully maps the HS model to a continuous one in any spatial dimension, \(d\). In principle, one can choose an arbitrary value for \(\lambda\), however, we follow the work of~\textcite{baez2018using} and set the value of \(\lambda=50\). For each spatial dimension, a value of the reduced temperature will be obtained. In particular, for three dimensions we use \(k_{B} T_{\mathrm{3D}} / \epsilon=1.4737\), and for four dimensions we use \(k_{B} T_{\mathrm{4D}} / \epsilon=1.4803\) ~\cite{baez2018using}.

The HS and pHS models have only one relevant parameter, which is the filling or packing fraction, defined for any spatial dimension as
\begin{equation}
    \eta = V_{d} \rho \sigma^{d} = \frac{\pi^{d/2}}{\Gamma{\left(1 + d/2 \right)}} \rho \sigma^{d}\; ,
    \label{eq:filling}
\end{equation}
with \(\rho\) being the number density, the hypersphere diameter is noted by \(\sigma\), and \(\Gamma{(x)}\) is the Gamma function.

In general, the compressibility factor in \(d\) dimensions can be written as follows~\cite{bishopEquationStateHard2005},
\begin{equation}
    Z = 1 + \frac{\rho}{2d}\int_0^\infty d\mathbf{r} \left(\frac{\partial \beta u}{\partial r}\right) g(r) r 
    \label{eq:virial_pressure} \, ,
\end{equation}
where \(d\mathbf{r}\) stands for the \(d\)-dimensional differential volume element. 
This expression is generic for any interaction potential \(u(r)\). However, for the HS interaction potential, the compressibility factor in \(d\) dimensions is related to the contact value of the pair correlation function \(g(\sigma^{+})\) by~\cite{bishopEquationStateHard2005}
\begin{equation}
    Z = \frac{\beta P}{\rho} = 1 + B_{2} \rho g(\sigma^{+}) \, ,
    \label{eq:z-eos}
\end{equation}
with \(Z\) the compressibility factor and \(\beta = 1 / k_{B} T\) is the inverse temperature. The second virial coefficient, \(B_{2}\), is defined for any spatial dimension \(d\) as~\cite{lubanThirdFourthVirial1982}
\begin{equation}
    B_{2} = \frac{\pi^{d/2} \, \sigma^{d}}{2 \, \Gamma(1 + d/2)} \, .
    \label{eq:b2}
\end{equation}
\section{\label{sec:methods} Computer simulation and integral equation theory details}
\subsection{\label{simulations} Computer simulations}
\subsubsection{Three dimensional simulations}
For the three dimensional system of particles interacting with the HS potential, we simulate a system of \(N=20376\) particles in a fixed volume \(V\), and energy \(E\). We perform event-driven molecular dynamics (EDMD) using the algorithm of~\textcite{smallenburgEfficientEventdrivenSimulations2022a}. Initial configurations are obtained by starting in a dilute state at the desired density, and then performing an EDMD simulation in which the particle diameters grow until the desired packing fraction is reached. After the packing fraction was reached, the system is equilibrated for at least \(10^{7} \tau_{\mathrm{MD}}\) and data collection for the pressure was done for another \(10^{8} \tau_{\mathrm{MD}}\), with \(\tau_{\mathrm{MD}}=\sigma \sqrt{m/k_{B} T}\) the simulation time unit, \(m\) the mass of the particle, which is the same for all particles. Pressure is measured using the virial expression~\cite{smallenburgEfficientEventdrivenSimulations2022a}. For the solid branch, all details remain the same except for the initial configuration, for which a face centered cubic (FCC) lattice was used for each of the target packing fractions. We estimate the average value of the pressure and the error of the measurement using block analysis~\cite{jonssonStandardErrorEstimation2018}.

The three dimensional pHS model was simulated using the HOOMD-blue code~\cite{andersonHOOMDbluePythonPackage2020}. Brownian dynamics (BD) simulations were performed using the standard Euler scheme to solve the equations of motion for the particles in the canonical (\(NVT\)) ensemble~\cite{ermakBrownianDynamicsHydrodynamic1978,snook2006langevin}. For the fluid branch of the equation of state, we simulate a total of \(N=32000\) particles at a fixed reduced temperature of \(k_{B} T / \epsilon= 1.4737\) following the result from~\textcite{baez2018using}. We use a time step of \(\Delta t = 10^{-5} \tau_{\mathrm{BD}}\), where \(\tau_{\mathrm{BD}} = \sigma^{2} / D_{0}\) is the Brownian time unit, and \(D_{0} = k_{B} T / 3 \pi \eta_{0} \sigma\) is the free-particle diffusion coefficient, with \(\eta_{0}\) being the zero-frequency shear viscosity. We initialize the system at a low density and slowly compress the simulation box over \(10^{5}\) time steps until the target packing fraction has been reached. Once the desired density has been reached, the system is equilibrated for at least \(5 \times 10^{7}\) time steps, and an additional production run of \(10^{7}\) time steps is used to collect data for the pressure, which is computed through the virial expression for a pairwise interaction potential~\cite{frenkel2023understanding}. For the solid branch of the equation of state of the pHS model, we use a smaller time step of \(\Delta t = 5 \times 10^{-7} \tau_{\mathrm{BD}}\), and initialize the system in the FCC configuration for the desired target packing fraction.

The reduced long-time self-diffusion coefficient, $D_{L}/D_{0}$, of hard spheres was computed from the linear fit of the mean-square displacement \cite{allen2017computer}, $W(t)\equiv \langle [\vec{r}(t)-\vec{r}(0)]^{2}\rangle$, at long times, where $\langle \cdots \rangle$ denotes an ensemble average of all particle trajectories and $\vec{r}(t)$ is the particle position at time $t$. In this case, and to save computational time, we consider colloidal dispersions made up of \(N=10976\) spherical particles. After an equilibration period of \(5 \times 10^{6}\) time steps, the production runs are carried out for \(10^{8}\) time steps to ensure a sufficiently large time window such that $W(t)$ reaches the linear diffusive regime~\cite{ermakBrownianDynamicsHydrodynamic1978}, i.e., $W(t)\sim 6 D_{L}t$ . We use the same reduced time step as for the fluid region defined previously.

To determine the fluid-crystal coexistence properties in the three dimensional pHS model, we calculate the coexistence pressure using the \(NVT\) ensemble method introduced by \textcite{smallenburgSimpleAccurateMethod2024a}. Below, we provide a brief summary of the method, but we direct readers to the original publication for a complete description of the method. The coexistence pressure in a fluid-crystal system is estimated by performing direct coexistence simulations in the \(NVT\) ensemble to evaluate the system's pressure tensor. The pressures of the fluid and solid are then compared, as true coexistence requires the pressures to be equal to satisfy mechanical equilibrium. For these simulations, we use \( N = 11,200 \) particles in an elongated simulation box oriented along the \( z \)-axis, and we create enough space in the simulation box to account for both phases, corresponding to a global packing fraction of \( \eta^{\mathrm{global}} = 0.5184 \). The initial particle configuration is arranged in a FCC lattice oriented with the square face perpendicular to the interface, with a crystal packing fraction in the range \( \eta^{X} \in [0.5475, 0.5575] \). Brownian dynamics simulations are performed for varying packing fractions, incremented by \( \Delta \eta^{X} = 10^{-4} \). Since the global packing fraction of the system is lower than the initial crystal fraction, the system undergoes phase separation. We simulate this phase separation for \( 5 \times 10^{7} \) time steps using the same time step as that employed for the solid branch of the equation of state. This equilibration period ensures the stabilization of the interfaces between the two coexisting phases. Additionally, we compute the liquid's equation of state within the range \( \eta \in [0.49, 0.5] \) in steps of \( \Delta \eta = 10^{-4} \). We then use the pressure tensor along the \( z \)-axis, \( P_{zz} \), obtained from the coexistence simulations and the equation of state of the liquid \( \beta P / \rho \) to calculate the coexistence pressure. The coexistence pressure is obtained by determining the point at which the pressures of the fluid and the fluid-crystal coexistence system are equal. To achieve this, we fit a straight line to the fluid's equation of state and a second-order polynomial to the pressure tensor results from the coexistence simulations, then apply a root-finding algorithm to solve \( \beta P_{zz} / \rho^{X} - \beta P / \rho = 0 \). To estimate the coexistence pressure and its uncertainty, we employ a bootstrapping method. Specifically, we randomly resample the data with replacement used for fitting both the linear and polynomial curves, generating \( n_{b} = 10,000 \) bootstrap samples. From these samples, we compute the mean and standard deviation, which are reported as the coexistence pressure and its associated uncertainty.

\subsubsection{Four dimensional simulations}
For the four dimensional hard-hypersphere model, we use EDMD simulations to obtain the equation of state only for the liquid branch. We use the implementation of~\textcite{skoge2006packing} and modify it to perform the simulations in this work. To create the initial configurations, we use the algorithm of~\textcite{skoge2006packing}, which is a modified Lubachevsky-Stillinger algorithm, to pack \(N=20376\) hyperspheres in a four dimensional simulation box until a target packing fraction was reached, and we ensure to have a small enough expansion rate such that the target packing fraction achieved has a relative difference of at least \(10^{-3}\) between the target and the computed packing fraction. After packing the simulation box, we equilibrate the system in the \(NVE\) ensemble for at least \(10^{7} \tau_{\mathrm{MD}}\); to collect data for the pressure of the system we simulate for an additional \(10^{8} \tau_{\mathrm{MD}}\).

To perform BD simulations of the four dimensional pHS model, we use an in-house BD code that implements the Ermak-McCammon algorithm~\cite{ermakBrownianDynamicsHydrodynamic1978} extended to four dimensions to solve the BD equations of motion. For the fluid branch of the equation of state, we simulate \(N=10000\) particles in the \(NVT\) ensemble with a fixed reduced temperature of \(k_{B} T / \epsilon = 1.4803\)~\cite{baez2018using}. Using a time step of \(\Delta t = 10^{-5} \tau_{\mathrm{BD}}\), we equilibrate the system for at least \(10^{5}\) time steps, and we collect data for the pressure during an additional simulation time of \(10^{6}\) time steps. For the solid branch of the equation of state, we initialize a \(D_{4}\) lattice~\cite{bishopEquationStateHard2005,vanmeelHardsphereCrystallizationGets2009} comprised of \(N=2048\) particles at the target packing fraction. We equilibrate the system for \(10^{6}\) time steps, and we collect data during an additional \(5 \times 10^{6}\) time steps; we use the same timestep as the one used for the fluid branch.

For all simulations in three or four dimensions, standard periodic boundary conditions were used in all directions of the simulation box. However, it is important to note that standard cubic periodic boundary conditions are not efficient or ideal in dimensions \(d > 3\) The conventional cubic periodic boundary condition, which corresponds to a simple cubic tiling, is clearly not optimal, because this sphere packing lattice is not the densest in any ~\cite{charbonneauDimensionalEvolutionStructure2022} \(d \geq 2\).
Furthermore, non-cubic or non-hypercubic simulation boxes can lead to an efficiency improvement for the computation of nearest neighbors, which helps to define neighbor lists for faster neighbor interaction computations.

\subsection{Integral equation theory}
One can obtain the molecular thermodynamic description of a classical liquid that is both uniform and isotropic by solving the Ornstein–Zernike (OZ) equation~\cite{hansen2013theory}
\begin{equation}
    h(r) = c(r) + \rho \int d\mathbf{r'} c (|\mathbf{r}-\mathbf{r'}|)h(\mathbf{r'}),
    \label{eq:oz}
\end{equation}
which serves as the definition of the direct correlation function~\cite{hansen2013theory} \(c(r)\). Here, \(h(r)=g(r)-1\) denotes the total correlation function, with \(g(r)\) being the radial distribution function. As noted, Eq.~\eqref{eq:oz} is a nonlinear integral equation, which typically must be solved numerically. One may exploit the convolution structure of the integral in Eq.~\eqref{eq:oz}, also known as the indirect correlation function \(\gamma (r) = h(r) - c(r)\), and turn this relation into an algebraic equation employing a Fourier transform (FT)~\cite{Ng10.1063/1.1682399},
\begin{equation}
    \hat{\gamma}(k) = \frac{\rho\hat{c}^2(k)}{1-\rho\hat{c}(k)} \, ,
\label{eq:fourier-transform}
\end{equation}
where \(k\) is the magnitude of the wave vector and the notation \(\hat{\gamma}(k)\) indicates that the function \(\gamma(r)\) is defined in the Fourier space.

However, a closure relation involving both $\gamma(r)$ and $u(r)$ is required. It is convenient to write it as follows~\cite{hansen2013theory},
\begin{equation}
    c(r) = \exp(-\beta u(r) + \gamma(r) + b(r)) -\gamma(r) - 1,
    \label{eq:second_condition}
\end{equation}
where \(b(r)\) is the so-called bridge function~\cite{hansen2013theory}. In general, \(b(r)\) is unknown and depends on the nature of the interaction potential. Although it is common to express \(b(r)\) as an infinite series without a closed-form expression, several closed approximations for \(b(r)\) have been developed and can be used to accurately study the thermodynamic properties of HS-like fluids.
Several approximations for the bridge function can be found in the literature~\cite{Bomont2008, solana2013perturbation}, however, in this work we focus on a modification to the semiphenomenological Verlet closure (MV) proposed by Kinoshita~\cite{kinoshita10.1063/1.1566935},
\begin{equation}
    b_{MV}(r) = - \frac{0.5 [\gamma(r)] ^2}{1 + 0.8 \lvert \gamma(r) \rvert} \, ,
    \label{eq:kinoshita}
\end{equation}
which has been recently used to account for the structure, thermodynamics, and depletion forces of binary mixtures of HS's even near thermodynamic instabilities; see, e.g.,~\cite{lopez2013demixing} and references therein.

The absolute value in Eq.~\eqref{eq:kinoshita} prevents divergence when the quantity \(1 + 0.8\gamma(r)\) approaches zero, thus improving the numerical stability of the approximation~\cite{bedolla_ev_opt}. A recent study~\cite{PhysRevE.110.044608Ilian} demonstrates that the MV closure more accurately reproduces the value of the radial distribution function at contact in a HS system compared to other bridge functions. However, to our knowledge, Eq.~\eqref{eq:kinoshita} has not been used to solve the OZ equation~\eqref{eq:oz} in four dimensions. Furthermore, the Percus–Yevick (PY) approximation, which provides an exact solution to the OZ equation in three dimensions~\cite{hansen2013theory}, has historically been used to compute the equation of state for HS-like fluids~\cite{barkerPerturbationTheoryEquation1967}. Therefore, to ensure completeness and to show the precision of $b_{MV}(r)$ when applied to both hard and pseudo-hard systems, we also utilized the PY approximation~\cite{hansen2013theory}, which is given by
\begin{equation}
    b_{PY}(r) = \ln[1 + \gamma (r)] - \gamma(r) \, .
    \label{eq:PY}
\end{equation}

To numerically solve the OZ equation, we used a recently developed software written in the Julia programming language\cite{Pihlajamaa2024, PhysRevE.110.044608Ilian}. This open-source code was designed to solve the OZ equation in arbitrary spatial dimensions and contains the necessary functions for calculating thermodynamic properties, such as the isothermal compressibility and the virial pressure.
We obtain the radial distribution function for different packing fraction values in order to construct the equation of state. The OZ integral equation is numerically solved at the cutoff radius of \(r_{c} = 10\sigma\), we use \(\mathrm{M}=2^{16}\) points within the interval of integration, and a grid spacing of \(\mathrm{dr}=r_{c}/\mathrm{M}\). To reach the target densities, a density ramp is employed, and a sequence of initial values is used to create a good initial estimate, this is the Ng iteration scheme~\cite{Ng10.1063/1.1682399}.

These numerical aspects become highly relevant since typical iterative algorithms require the computation of the radially symmetric \(d\)-dimensional Fourier transform of the direct correlation function, which is related to the Hankel transform~\cite{fourierND} and is given by~\cite{PhysRevE.110.044608Ilian}
\begin{equation}
    \hat{f}(k) = \frac{(2\pi)^{d/2}}{k^{d/2-1}}\int_0^\infty  dr J_{d/2-1} f(r) r^{d/2} \, ,
    \label{eq:ft}
\end{equation}
where \(f(r)\) is an arbitrary radial function and \(J_m(x)\) represents the Bessel function of the first kind of order \(m\). Once we calculate the FT of Eq.~\eqref{eq:second_condition} using an initial guess for \(\gamma(r)\), we use it as input to build new guesses with the aid of \(\hat{\gamma}(k)\). Then, we return to real space applying the inverse Fourier transform~\cite{Ng10.1063/1.1682399}.
One should be aware that dealing with even spatial dimensions (such as in our four-dimensional case) entails the appearance of Bessel functions of integer order. In contrast, working in three dimensions leads to the more common Fourier-Sine transform due to the properties of half-integer Bessel functions~\cite{Arfken2013-dr}.

Once we find the solution of the OZ equation, the corresponding \(g(r)\) is used to construct the equation of state using Eq.~\eqref{eq:virial_pressure}.
The latter involves an improper integral of the product of the \(g(r)\) and the radial derivative of the continuous potential \(u_{pHS}(r)\).
Since the potential decays rapidly at contact and quickly approaches zero at larger distances, and because \(g(r)\) becomes nonzero only just before contact, the integration window in Eq.~\eqref{eq:virial_pressure} is very narrow. Consequently, the grid used to solve the OZ equation must be sufficiently fine, containing enough points to accurately perform the numerical integration required to compute the compressibility factor via Eq.~\eqref{eq:virial_pressure}.

\section{\label{sec:integral}Fluid regime: Integral equation predictions and mean-field approximations}
In this section we show the results of solving the OZ equation using the the MV closure, Eq.~\eqref{eq:kinoshita} and the PY closure, Eq.~\eqref{eq:PY}. The solutions are compared directly to the simulation results obtained for the BD simulations of the pHS model and the EDMD simulations of the HS model, in both $3D$ and $4D$. Additionally, we compare the results for the latter with a well-known mean-field approach.

\subsection{Three dimensional equation of state}

\begin{figure}
    \includegraphics[width=0.85\columnwidth]{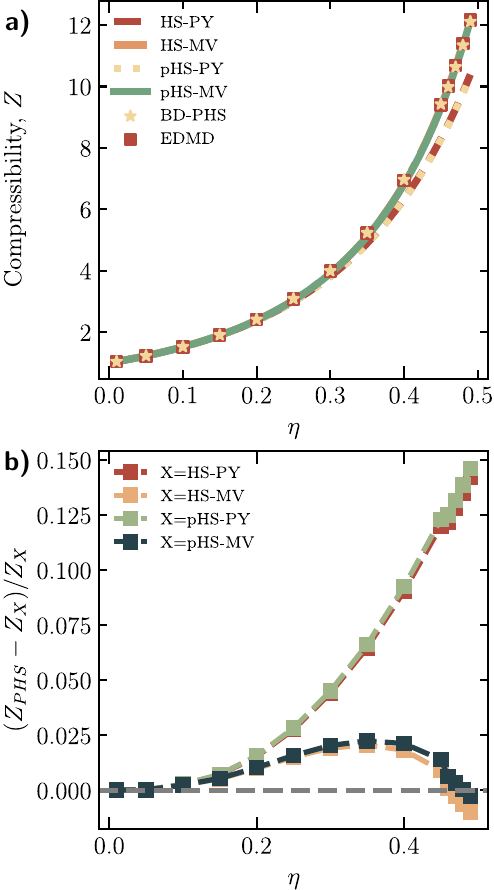}
    \caption[3doz]{\textbf{Equation of state for the three dimensional pseudo hard-sphere fluid from the Ornstein-Zernike equation for different closure relations.}
    \begin{enumerate*}[label=\textbf{\alph*})]
         \item The compressibility factor \(Z=\beta P / \rho\) as a function of the packing fraction \(\eta = \pi \rho \sigma^{3} / 6\) for the pseudo hard sphere fluid in three dimensions as obtained with Brownian Dynamics simulations (BD-PHS), and for the real hard sphere fluid as obtained with event-driven molecular dynamics (EDMD). We solve the Ornstein-Zernike relation for two closure relations, namely, the Percus-Yevick (PY) and the Kinoshita modification to the Verlet expression (MV). The labels HS and pHS represent the interaction potential used to solve the Ornstein-Zernike equation.
         \item Relative deviation \((Z_{PHS} - Z_{X}) / Z_{x}\) as a function of the packing fraction \(\eta\) for the liquid branch. The labels are the same as in panel a).
         \end{enumerate*}}
    \label{fig:oz-3d}
\end{figure}

In Fig.~\ref{fig:oz-3d}, the equation of state found from solving the OZ is presented, along with the simulation results for both the HS and pHS interaction potentials.

From Fig.~\ref{fig:oz-3d}a), we observe that the PY closure is not precise enough to reproduce the compressibility factor at high filling fractions \(\eta > 0.4\), which has been a well-known fact from liquid state theory~\cite{hansen2013theory}. Furthermore, the difference between the HS and pHS models is not noticeable, alluding to the fact that both models are of comparable precision in the values obtained for the compressibility factor, something also observed in the simulations results.

On the other hand, the Kinoshita variation of the Verlet closure is precise enough even for the high filling fractions, being able to reproduce the simulation results quite well, as seen in Fig.~\ref{fig:oz-3d}a), for both the HS and the pHS models. The Verlet closure has been closely examined and extensively tested for the HS model~\cite{kinoshita10.1063/1.1566935,lopez2013demixing,perera2016assessment,bedolla_ev_opt,PhysRevE.110.044608Ilian}. It has been shown that this closure can accurately reproduce the thermodynamic properties of the HS model and also the pHS model.

The deviation in the compressibility factor shows a clearer picture on the precision of the closures for the OZ equation by comparing it to the simulation results from BD simulations. In Fig.~\ref{fig:oz-3d}b), we observe that the deviations start to increase consistently starting for the PY closure from a small filling fraction of about \(\eta \approx 0.2\), for both the HS and pHS models, indicating what we saw before that the PY is not a good closure for the HS model or even the pHS model. This is well-established, but it is good to see that we can reproduce here as well, even for the pHS model. On the other hand, the MV closure is highly accurate and reproduces the compressibility factor from simulations quite well, with the HS model having a smaller error in the measurement. This goes on to show that the closure remains a good approximation for solving the HS model, or similar interaction potentials, and it is a good idea to use this as a first attempt at solving the OZ equation for similar interaction potentials if no other knowledge of the closure relation is known beforehand.

\subsection{Four dimensional equation of state}
\begin{figure}
    \includegraphics[width=0.85\columnwidth]{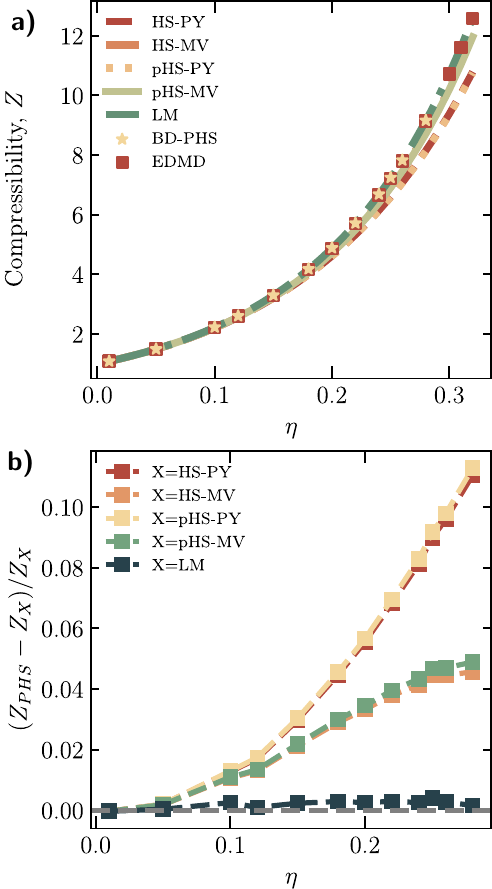}
    \caption[4doz]{\textbf{Equation of state for the four dimensional pseudo hard-sphere fluid from the Ornstein-Zernike equation for different closure relations.}
    \begin{enumerate*}[label=\textbf{\alph*})]
         \item The compressibility factor \(Z=\beta P / \rho\) as a function of the packing fraction \(\eta = \pi^2 \rho \, \sigma^{4} / 32\) for the pseudo hard sphere fluid in four dimensions as obtained with Brownian Dynamics simulations (BD-PHS), and for the real hard hypersphere fluid as obtained with event-driven molecular dynamics (EDMD). We solve the Ornstein-Zernike relation for two closure relations, namely, the Percus-Yevick (PY) and the Kinoshita modification to the Verlet expression (MV). The labels HS and pHS represent the interaction potential used to solve the Ornstein-Zernike equation. The empirical Luban-Michels (LM) defined in Eq.~\eqref{Zlm} is also shown.
         \item Relative deviation \((Z_{PHS} - Z_{X}) / Z_{x}\) as a function of the packing fraction \(\eta\) for the liquid branch. The labels are the same as in panel a).
         \end{enumerate*}}
    \label{fig:oz-4d}
\end{figure}

Now we turn our attention to the four dimensional case and the solution of the OZ equation, shown in Fig.~\ref{fig:oz-4d}. The case of four dimensions is interesting because there is a scarcity in the literature on this topic, although the topic of higher dimensional HS models is of relevance, as has been pointed out previously. In particular, equations of state for mixtures of hard hyperspheres have been proposed and analyzed in four and five dimensions~\cite{santosEquationStateMulticomponent1999, santosVirialCoefficientsEquations2001,10.1063/1.1349094LH, S.B.Yuste_2000,e22040469}. These mean-field approximations are written in terms of the equation of state of the one-component hard-hypersphere fluid. Therefore, an accurate expression for the compressibility factor of the one-component fluid is needed. A mean-field equation of state of hard \(d\)-dimensional hyperspheres is defined by~\textcite{Luban1990} as follows, 
\begin{equation}
Z_{\mathrm{LM}}(\eta) = 1 + b_{2} \, \eta
\frac{
  1 
  + \left[ \frac{b_{3}}{b_{2}} 
       - \zeta(\eta) \frac{b_{4}}{b_{3}} \right] \eta
}{
  1 
  - \zeta(\eta) \left( \frac{b_{4}}{b_{3}} \right)\eta
  + \left[ \zeta(\eta) - 1 \right] \left( \frac{b_{4}}{b_{2}} \right) \eta^{2}
} \, ,
\label{Zlm}
\end{equation}
which incorporates the exact expressions for the reduced virial coefficients $b_2$, $b_3$ and $b_4.$ The $k$-th reduced virial coefficient can be written in terms of $V_d$,~\cite{Santos2016_chap3}
\begin{equation}
   b_k = \left( \frac{V_d}{2^d}\sigma^d \right)^{-(k-1)} B_k \, ,
    \label{eq:reduced_virial_coefficient}
\end{equation}
with \(B_k\) being the \(k\)-th virial coefficient.
The coefficients of the linear function \(\zeta(\eta) = \zeta_0 + \zeta_1 (\eta/\eta_{cp})\), with \(\eta_{cp}\) representing the crystalline close-packing value, are obtained using computer simulations and the known virial coefficients.
All quantities in Eq.~\eqref{Zlm} can be found explicitly in Table~\ref{tab:params}.

\begin{table}[h]
\centering
\begin{tabular}{ll}
\hline
\textbf{Parameter} & \textbf{Value} \\
\hline
$b_{2}$            
  & $8$ \\[5pt]
$b_{3}$            
  & $2^{6}\!\Bigl(\tfrac{4}{3} - \tfrac{3\sqrt{3}}{2\pi}\Bigr)$ \\[5pt]
$b_{4}$            
  & $2^{9}\!\Bigl(2 \;-\; \tfrac{27\sqrt{3}}{4\pi} \, + \, \tfrac{832}{45\pi^{2}}\Bigr)$ \\[5pt]
$\zeta_{0}$        
  & $1.2973(59)$ \\[5pt]
$\zeta_{1}$        
  & $-0.062(13)$ \\[5pt]
$\eta_{\mathrm{cp}}$ 
  & $\frac{\pi^{2}}{16}$ \\
\hline
\end{tabular}
\caption{Reduced virial coefficients and parameters of the Luban-Michels equation of state~\eqref{Zlm} for \(d=4\) obtained from L{\'o}pez de Haro \emph{et al.}~\cite{e22040469}}.
\label{tab:params}
\end{table}

Panel (a) in Fig.~\ref{fig:oz-4d} shows that the Luban–Michels (LM) equation of state, defined in Eq.~\eqref{Zlm}, exhibits excellent accuracy compared to simulation data. This remarkable agreement arises from the theoretical formulation of the equation as a ratio of polynomials that incorporates the virial coefficients, thereby yielding an accurate representation of the virial expansion across all densities up to the liquid–solid transition. Notably, it is surprising that the original work~\cite{lubanThirdFourthVirial1982}, which was based on a relatively small system of approximately \(N=684\) particles, achieved sufficient precision to reproduce the compressibility factor observed in simulations of the much larger systems simulated in this work. This suggests that even with higher-accuracy data, further improvements in the predictive capability of the LM equation of state for the pressure of the four-dimensional HS fluid would likely be marginal.

Figure~\ref{fig:oz-4d}a) shows that the PY closure performs significantly worse at higher filling fractions \(\eta \geq 0.15\), where the compressibility factor begins to deviate from simulation results. In Figure~\ref{fig:oz-4d}b, deviations are observed starting at approximately \(\eta \geq 0.15\), and they increase steadily with increasing filling fraction until a maximum deviation is reached, corresponding to the fluid-solid transition (data not shown). The fluid-solid transition is known to be challenging for the OZ equation due to the numerical instability of conventional algorithms. Thus, the PY closure is less reliable as an initial approach for obtaining thermodynamic properties of higher-dimensional HS-like fluids.

On the other hand, the MV closure again exhibits good accuracy, as shown in Fig.~\ref{fig:oz-4d}a). However, at high densities, around \(\eta \approx 0.2\), it fails to provide accurate estimates of the compressibility factor. This discrepancy is evident in Fig.~\ref{fig:oz-4d}b), where the deviations increase with the filling fraction. Compared to the three-dimensional fluid, the deviations in the current case are larger and exhibit a steeper increase with density. This observation suggests that the original Verlet closure, defined only for the three-dimensional HS fluid~\cite{verlet1981integral}, might require modifications in its functional form to properly account for the virial coefficients in higher dimensions. A procedure similar to that employed in~\textcite{bedolla_ev_opt}, which utilizes a machine learning framework to obtain the coefficients for a generalized functional MV closure, could be beneficial. Nevertheless, the MV closure demonstrates sufficient accuracy to reproduce the compressibility factor values observed in simulations. This promising result underscores the robustness of the MV closure in contexts beyond its original intended application. Future investigations should examine whether this closure remains effective in higher dimensions and determine its corresponding accuracy in those cases.

\section{\label{sec:diff}Long-time self-diffusion coefficient of the pseudo hard-sphere potential}

Although this contribution mainly deals with the equilibrium phase diagram and, therefore, thermodynamic properties are determined, we consider it relevant to analyze and discuss whether the particle trajectories obtained from the Brownian dynamics (BD) simulations of the pHS potential correctly reproduce the transport properties of hard spheres in the so-called diffusive regime. This case corresponds to the dynamics of a colloidal dispersion made up of hard spheres \cite{medina-noyolaLongTimeSelfDiffusionConcentrated1988}. However, as we discuss below, there exists a heuristic criterion based on the long-time self-diffusion coefficient that also provides an estimation of the liquid-solid transition \cite{lowenDynamicalCriterionFreezing1993}. This route will be an independent confirmation that the pHS model reproduces not only the thermodynamic properties of hard spheres but also the transport phenomena.

\begin{figure}
    \includegraphics{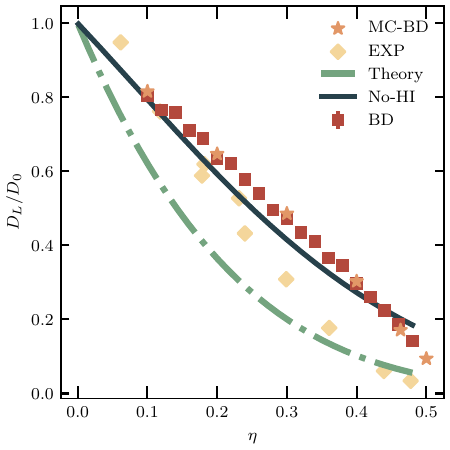}
    \caption{The reduced long-time self-diffusion coefficient, \(D_{L} / D_{0}\), as a function of the packing fraction in three dimensions, \(\eta = \pi \rho \sigma^{3} / 6\), is shown. The Brownian dynamics simulations of the pHS potential are shown (close square) with error bars smaller than than the symbol size. The results are compared against experiments (close diamonds) of colloidal suspensions (EXP)~\cite{vanblaaderenLongtimeSelfdiffusionSpherical1992} and dynamic Monte Carlo simulations (closed stars) of hard spheres without hydrodynamic interactions~\cite{cichockiDynamicComputerSimulation1990} (stars; MC-BD). Also, theoretical predictions from \textcite{medina-noyolaLongTimeSelfDiffusionConcentrated1988} (lines; Theory) with and without HI are also displayed.}
    \label{fig:diffusion}
\end{figure}

As explained above, we have extracted the ratio $D_{L}/D_{0}$ from the mean-square displacement. The latter is computed using the pHS potential for every packing fraction. Then, in Fig.~\ref{fig:diffusion}, we present the BD simulation results without hydrodynamic interactions (HI), indicated by square symbols. These results are compared with experimental measurements of colloidal suspensions~\cite{vanblaaderenLongtimeSelfdiffusionSpherical1992} nearly interacting as hard spheres, labeled as EXP. We also include results from dynamic Monte Carlo simulations of hard spheres without HI \cite{cichockiDynamicComputerSimulation1990}, as well as the theoretical predictions of Medina-
Noyola with and without HI \cite{medina-noyolaLongTimeSelfDiffusionConcentrated1988}; this dynamical approach made use of the Percus-Yevick approximation as the static input. We first compare the predictions of the pHS potential with the experiments. As seen, the agreement is good at low volume fractions (\(\eta < 0.2\)); however, the discrepancies become more pronounced at higher densities. We attribute this deviation to the presence of HI in the experimental systems. In fact, the original experimental work includes a comparison with the simulation data that accounts for HI~\cite{cichockiDynamicComputerSimulation1990}, highlighting significant differences at higher densities. Interestingly, the theoretical framework including HI seems to perform well at high concentrations, but the case without HI nicely follows the predictions from the pHS, which basically reproduce the dynamic Monte Carlo data for hard spheres. This level of agreement allows us to conclude that the Brownian dynamics of the pHS potential correctly reproduces the diffusive behavior of hard spheres without HI. Nonetheless, results from quasi-two-dimensional hard-disk experiments~\cite{thorneyworkEffectHydrodynamicInteractions2015} suggest that HI have a negligible impact on the long-time self-diffusion coefficient, with good agreement observed between experimental and simulation data. This raises the question of whether a similar dynamical behavior might hold in three-dimensional systems. We propose that a renewed experimental effort to measure the long-time self-diffusion coefficient together with the use of the pHS potential in Brownian dynamics simulations including explicitly HI could help resolve this discrepancy and clarify the role of HI in dense colloidal suspensions; work along this line is in progress.

Furthermore, we note that the BD results with the pHS potential also show excellent agreement with the Lowen's freezing criterion\cite{lowenDynamicalCriterionFreezing1993}, since the value of the long-time self-diffusion coefficient at a packing fraction of \(\eta=0.48\) is \(D_{L}/D_{0}=0.143 \pm 0.001\), which is close to the predicted value of \(D_{L}/D_{0}=0.1\) and this case corresponds to the liquid-solid transition point for hard spheres, \(\eta=0.49\), which we will study in detail in the following section.
\\

\section{\label{sec:sims}Phase diagram of the pseudo hard-sphere potential}

\subsection{Three dimensional equation of state}
\begin{figure}
    \includegraphics[width=0.9\columnwidth]{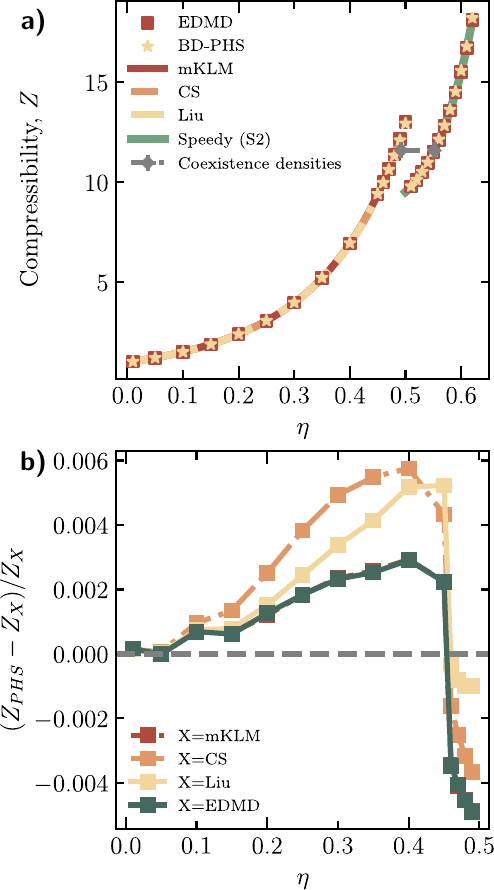}
    \caption[3dsim]{\textbf{Phase diagram of the pseudo hard sphere potential in three dimensions.}
    \begin{enumerate*}[label=\textbf{\alph*})]
         \item The compressibility factor \(Z=\beta P / \rho\) as a function of the packing fraction \(\eta = \pi \rho \sigma^{3} / 6\) for the pseudo hard sphere fluid as obtained with Brownian Dynamics simulations (BD-PHS), and for the real hard sphere fluid as obtained with event-driven molecular dynamics (EDMD). We also compare to mean-field equations of state, such as the Carnahan-Starling equation of state (CS)~\cite{hansen2013theory} and the equation of state by Liu~\cite{liuCarnahanStarlingTypeEquations2021b} that follows the CS approach for all available virial coefficients of the hard sphere fluid. We also compare to the highly accurate equation of state that follows a virial expansion fitted with simulation data (mKLM)~\cite{pieprzykThermodynamicDynamicalProperties2019a} for the fluid, and the reparametrization S2 using EDMD simulation data of the Speedy equation of state for the solid branch.~\cite{speedyPressureEntropyHardsphere1998,pieprzykThermodynamicDynamicalProperties2019a} \label{eos3d1}
         \item Relative deviation \((Z_{PHS} - Z_{X}) / Z_{x}\) as a function of the packing fraction \(\eta\) for the liquid branch. The equations of state are those as in~\ref{eos3d1}, as well as the simulation data obtained with EDMD. The gray dashed line is a guide to the eye.
         \end{enumerate*}}
    \label{fig:eos3d}
\end{figure}

We present our results for the equation of state of the three dimensional pHS and the HS model. For the fluid branch, we simulate densities in the range of \(\eta \in [0.01, 0.49]\), and for the solid branch, we simulate densities in the range \(\eta \in [0.5,0.62]\), just below the value of the densest packing \(\eta_{CP} = \pi / \sqrt{18}\). 
In Fig.~\ref{fig:eos3d}a) we present the results obtained from simulations and we compare it with several mean-field equations of state, as well as those fitted with simulation data. We use the well-known Carnahan-Starling (CS) equation of state~\cite{hansen2013theory}, and a re-fitted version of this equation of state using higher order virial coefficient due to~\textcite{liuCarnahanStarlingTypeEquations2021b}. For the simulation parametrized equations of state, we use the virial coefficient fits from~\cite{pieprzykThermodynamicDynamicalProperties2019a} for both the fluid and solid branches.

Figure~\ref{fig:eos3d}a) demonstrates that the pHS model accurately reproduces the qualitative behavior of the compressibility factor for the HS model, as well as the trends observed in mean-field equations of state. This outcome is consistent with prior findings, as the pHS model is well-known for its ability to replicate the equation of state~\cite{baez2018using} and the structural properties of the HS system~\cite{martinez-riveraModelingStructureThermodynamics2023,bedolla-montielInverseDesignCrystals2024}. 
To quantitatively assess the accuracy of the pHS model, Figure~\ref{fig:eos3d}b) presents the relative difference between the BD simulation results of the pHS model and various equations of state as a function of packing fraction.
At higher packing fractions, the difference between the pHS and Carnahan-Starling (CS) equations of state is notably larger, which is expected since the CS equation of state is known to exhibit reduced precision in this regime.
By contrast, the Liu equation of state aligns more closely with the simulation data, a result attributed to its inclusion of higher-order virial coefficients, which improve the accuracy of its estimation of the compressibility factor. The results obtained from EDMD simulations and the mKLM equation of state are nearly indistinguishable, as the mKLM equation is directly derived from EDMD simulation data. Consequently, they are compared equivalently, revealing that the pHS model performs well overall, except at high densities within the range \(\eta \in [0.4, 0.5]\). However, the relative difference in this region is smaller than previously reported~\cite{baez2018using}, a result we attribute to the use of a larger number of particles, which improves the accuracy of pressure measurements.

The coexistence pressure of the pHS model was measured to be \(\beta P \sigma^{3} = 11.66 \pm 0.04\), a result that is in excellent agreement with the previously reported value for the pHS model by~\textcite{joverPseudoHardspherePotential2012}, \(\beta P \sigma^{3} = 11.65 \pm 0.01\)~\cite{espinosaFluidsolidDirectCoexistence2013}. In the study by~\textcite{espinosaFluidsolidDirectCoexistence2013}, \(NPT\) simulations were employed to determine the coexistence pressure of the pHS model at a reduced temperature of \(k_{B} T / \epsilon = 1.5\), following the methodology of~\textcite{joverPseudoHardspherePotential2012}. These results confirm the suitability of the pHS model for determining coexistence properties and possibly investigating crystal nucleation. Notably, both measurements are consistent with the coexistence pressure of the HS model, reported as \(\beta P \sigma^{3} = 11.5645 \pm 0.0005\)~\cite{smallenburgSimpleAccurateMethod2024a}. However, given the relatively higher uncertainty in the pHS measurement, it remains an open question whether the free energy difference between the face-centered cubic (FCC) lattice and the hexagonal close-packed (HCP) structure~\cite{frenkel2023understanding} can be distinguished using the pHS model.

\subsection{Four dimensional equation of state}
\begin{figure}
    \includegraphics[width=0.9\columnwidth]{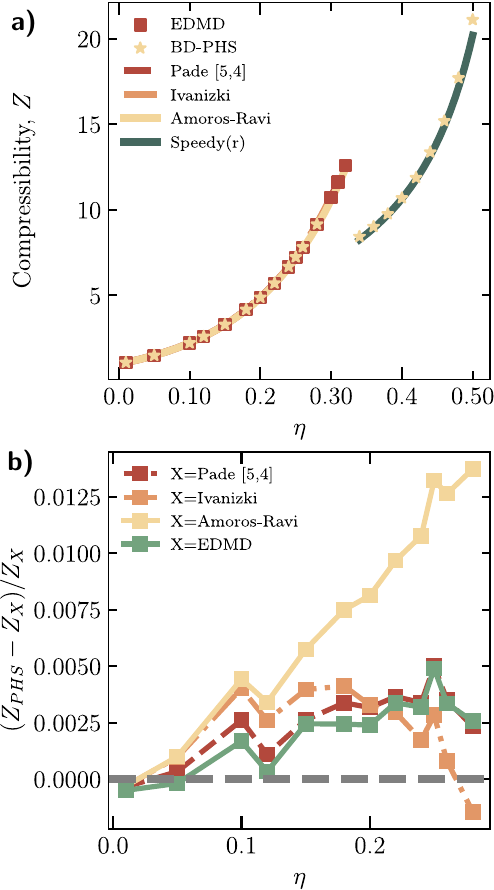}
    \caption[4dsim]{\textbf{Phase diagram of the pseudo hard sphere potential in four dimensions.}
    \begin{enumerate*}[label=\textbf{\alph*})]
         \item The compressibility factor \(Z=\beta P / \rho\) as a function of the packing fraction \(\eta = \pi^{2} \rho \sigma^{4} / 32\) for the pseudo hard sphere fluid as obtained with Brownian Dynamics simulations (BD-PHS), and for the real hard sphere fluid as obtained with event-driven molecular dynamics (EDMD). We also show results from mean-field equations of state, such as the expression by Ivanizki~\cite{IVANIZKI20181745} and the equation of state from Amor{\'o}s and Ravi~\cite{AMOROS20132089}. We also compare with the equation of state from~\textcite{bishopEquationStateHard2005}(Pad{\'e}[5,4]), which is fitted from Monte Carlo simulations. For the solid branch, the \(D_{4}\) crystal is used for the simulations, and we compare with the reparametrized Speedy expression (Speedy(r)) from~\textcite{lueFluidSolidPhase2010}. \label{eos4d1}
         \item Relative deviation \((Z_{PHS} - Z_{X}) / Z_{x}\) as a function of the packing fraction \(\eta\) for the liquid branch. The equations of state are those as in~\ref{eos4d1}, as well as the simulation data obtained with EDMD. The gray dashed line is a guide to the eye.
         \end{enumerate*}}
    \label{fig:eos4d}
\end{figure}

We now present the results for the four-dimensional pHS and hard-hypersphere models. Due to limitations in extending the methodology outlined in~\textcite{smallenburgSimpleAccurateMethod2024a} to compute the coexistence pressure for four dimensions, such results are not included in this work. Instead, we focus on presenting the equation of state obtained from BD simulations and comparing these results with EDMD simulation data and mean-field equations of state. The main findings are summarized in Fig.~\ref{fig:eos4d}, where panel Fig.~\ref{fig:eos4d}a) shows the results obtained for the equations of state. We use the Pad{\'e}[5,4] approximation~\cite{bishopEquationStateHard2005} derived from simulations of the four-dimensional hard-hypersphere fluid to compare with our BD simulations. We also compare with the Ivanizki equation of state~\cite{IVANIZKI20181745}, which applies a generalized CS approach based on the available virial coefficients for the four-dimensional hard-hypersphere fluid. Similarly, the equation of state by Amor{\'o}s and Ravi (AR) follows the CS approach, but does not constrain the polynomial representation of the virial coefficients to simulation data~\cite{AMOROS20132089}. Finally, for comparison with the solid branch, we utilize the semi-phenomenological equation of state by Speedy, which has been re-parameterized with simulation data for a \(D_{4}\) lattice~\cite{lueFluidSolidPhase2010}.

First, we note that the simulation data for the pHS model and the EDMD simulation results for the fluid branch are in excellent agreement, as shown in Fig.~\ref{fig:eos4d}a), for the packing fraction range \(\eta \in [0.01, 0.28]\). The relative deviation between these results is minimal, as illustrated in Fig.~\ref{fig:eos4d}b). The other mean-field equations of state also show strong agreement with one another, and the pHS model demonstrates good consistency with these equations of state. However, examining the relative deviation for both the Ivanizki and AR equations of state reveals that, at high packing fractions, the AR equation diverges significantly, with a substantial loss in precision. This discrepancy may stem from the AR equation's use of a pole with an order less than the dimensionality \(d\), and the imposition of integer coefficients to reproduce the fifth virial coefficient of the four-dimensional hard-hypersphere fluid appears unrelated to its ability to accurately predict the compressibility factor~\cite{IVANIZKI20181745}. Similar to the three-dimensional case, the EDMD simulation results and the Pad{\'e}[5,4] approximation~\cite{bishopEquationStateHard2005} exhibit nearly identical accuracy, which we attribute to the use of simulation data to fit the corresponding Pad{\'e} polynomials. Both approaches display comparable precision and underscore the high accuracy of the pHS model in reproducing results from these equations of state. In the final comparison for the fluid branch, the Ivanizki equation of state~\cite{IVANIZKI20181745} achieves a precision level similar to that of EDMD and Pad{\'e}[5,4], and the pHS BD simulations reproduce the results of the Ivanizki approximation with high fidelity. We regard the Ivanizki equation of state for four dimensional fluids as an excellent general-purpose expression, serving as a reliable analogue to the CS equation of state for three-dimensional fluids.

The results for the solid branch of the four-dimensional pHS model are presented in Fig.~\ref{fig:eos4d}a) for the packing fraction range \(\eta \in [0.34, 0.5]\). These results are compared with the Speedy re-parametrization~\cite{lueFluidSolidPhase2010}, and we observe good agreement, except at higher packing fractions. At high packing fractions, the computed compressibility factor deviates significantly from the values predicted by the Speedy equation of state. A closer analysis reveals that the highest packing fraction simulated, \(\eta=0.5\), exceeds the maximally random jammed density of \(\eta = 0.46 \pm 0.005\)~\cite{skoge2006packing}, but remains below the densest packing density of \(\eta = 0.6169\)~\cite{conway2013sphere}. However, the results from Ref.~\cite{lueFluidSolidPhase2010} extend to packing fractions of at least \(\eta = 0.55\) and use this data to fit the Speedy equation of state. This discrepancy in precision may arise from the use of a small number of particles, which is a known limitation in high-dimensional systems and impacts the accuracy of thermodynamic measurements~\cite{skoge2006packing,charbonneauGlassTransitionRandom2011,charbonneauGeometricalFrustrationStatic2013,charbonneauDimensionalEvolutionStructure2022}. While the results exhibit qualitative agreement with the equation of state, improved BD simulations employing more efficient periodic boundary conditions~\cite{charbonneauDimensionalEvolutionStructure2022} would likely improve the precision of the compressibility factor in this regime.

\section{\label{conclusions}Concluding remarks and perspectives}
In this work, we have used computer simulations and integral equation theory to systematically study the phase diagram and thermodynamic properties of the pHS model introduced by~\cite{baez2018using}. 

First, we set out to solve the OZ equation for two specific closure relations, the Percus-Yevick and the modified Verlet approximation. We found that in both cases the PY closure is only accurate for low densities and when the density increases, the accuracy of the closure decreases considerably. On the other hand, for the three dimensional case, the MV closure relation correctly reproduced the compressibility factor of the BD simulation results, and the deviations between measurements were small, showing the high precision of the closure. For the four dimensional case, the MV closure showed less precision at higher densities, which we believe is due to the fact that the MV closure was originally described for a three dimensional fluid. In contrast, the empirical equation of state 
LM~\cite{lubanThirdFourthVirial1982,santosEquationStateMulticomponent1999} was shown to offer highly accurate results compared to the simulation results. This equation of state was formulated to use the virial coefficients available to the date of the original work, and computer simulation data of the four-dimensional HS fluid, providing enough data to fix the free parameters in the formulation.

Next, we employed BD simulations to obtain the compressibility factor of three and four dimensional fluids using the pHS model. For the particular case of the three dimensional fluid, we also simulated the solid branch of the system, obtaining a FCC crystal and computing the coexistence densities using a recent \(NVT\) direct coexistence method~\cite{smallenburgSimpleAccurateMethod2024a}. We also computed the equation of state of the true HS fluid in three and four dimensions using EDMD simulations. We compared these results to well-known mean-field equations of state, as well as other semi-phenomenological approximations, and simulations fits. We arrived at the conclusion that the pHS model can correctly reproduce the equation of state of both the three and the four dimensional fluids for all densities investigated. We also observed a good agreement with the coexistence density, and the solid branch of the three dimensional HS fluid agrees with previously reported data, our own EDMD simulation data, and equations of state.

A main conclusion of this work is the high accuracy of the pHS model in reproducing thermodynamic properties in three and four dimensions, something that was not tested before; in the original work only two and three dimensions were tested with much fewer particles than the ones employed in this contribution and within the fluid regime~\cite{baez2018using}. Higher dimensions are of interest to studies of the glass transition, and a model such as the pHS model enables the use of standard simulation techniques and continuous integration schemes, as well as optimizations well-known in the computer simulation field. Furthermore, this shows that the model can reproduce the properties of higher dimensional fluids, but the solid branches are still something that has to be studied in detail, since in higher dimensions different types of crystals are more stable than others.

Regarding the integral equation theory, the OZ equation has been known to be solvable in any dimension, and with current solvers, higher dimensions are possible, as shown in this work. However, there seems to be almost no literature on closure relations for higher-dimensional fluids, and with this work we want to shed light into the possibility of extending the well-known closure relation, like the MV approximation, to higher dimensional fluids. Machine learning frameworks might help improve the closure relations and provide more accurate results as well.

It is also of interest to investigate the role of hydrodynamic interactions (HI) in systems of pseudo-hard-spheres and pseudo-hard-hyperspheres, particularly regarding their influence on the long-time self-diffusion coefficient. While experimental data are available for hard-sphere systems, for higher-dimensional analogues, validation is limited to comparisons with theoretical predictions or alternative simulation techniques. In this work, we have shown that the interaction potential defined in Eq.~\eqref{eq:wca} can reproduce the dynamics of hard-sphere systems, closely approaching the freezing criterion predicted by L{\"o}wen \emph{et al.}~\cite{lowenDynamicalCriterionFreezing1993}. However, discrepancies with the experimental data are observed, which we attribute to the potential influence of HI as well as to the dated nature of the experimental results. A more systematic and comprehensive comparison between theory, simulations, and experiments appears feasible and would be valuable in assessing the extent to which the pHS model accurately captures the essential features of the dynamics and transport phenomena hard-sphere dynamics.

In closing, we have shown that the pHS model is robust enough to reproduce the thermodynamic properties of the established HS model, and with the aid of computer simulations and integral equation theory, we demonstrated that this model is an excellent candidate for modeling hard-like interactions in soft matter systems. It will be of interest to further test the generality of pHS in binary mixtures and polydisperse systems, since there exist mean-field approximations for the equations of state of these systems~\cite{santosEquationStateMulticomponent1999,santosVirialCoefficientsEquations2001}.

\begin{acknowledgments}
The authors acknowledge financial support from SECIHTI (Grant No. CBF2023-2024-3350) and the Marcos Moshinsky Foundation.
\end{acknowledgments}

\section*{Data Availability}
The data that support the findings of this study are openly available in the Zenodo repository \url{https://doi.org/10.5281/zenodo.15105653}.

\bibliography{bibliography}

\end{document}